\documentclass{ws-procs9x6}

\begin{document}

\title{Photoproduction of $\Theta^+$ on the nucleon and deuteron}

\author{T. Mart}

\address{Departemen Fisika, FMIPA, Universitas Indonesia, Depok 16424, Indonesia}

\author{A. Salam and K. Miyagawa}

\address{Department of Applied Physics, Okayama University of Science, 
  1-1 Ridai-cho, Okayama 700, Japan}
\author{C. Bennhold}

\address{Center for Nuclear Studies, Department of Physics, The George 
         Washington University, Washington, D.C. 20052, USA}  

\maketitle

\abstracts{
Photoproduction of the pentaquark particle $\Theta^+$ on the nucleon  
has been studied by using an isobar and a Regge model. Using the isobar model, 
total cross sections around 100 nb for the $\gamma n\to K^-\Theta^+$ 
channel and 400 nb for the $\gamma p\to {\bar K}^0 \Theta^+$ process are obtained.
The inclusion of the $K^*$ intermediate state  yields a substantially 
large effect, especially in the $\gamma p\to {\bar K}^0 \Theta^+$ 
process. The Regge approach predicts smaller cross sections, i.e., less than
100 nb (20 nb) for the process on the neutron (proton). By using an elementary
operator from the isobar model, 
cross sections for the process on a deuteron are predicted.
}

\section{Introduction}
The observation of the pentaquark $\Theta^+$ baryon\cite{experiment} has
triggered a great number of investigations on the production process of
this unconventional particle. In general, these efforts can be divided into two
categories, i.e., investigations using hadronic and electromagnetic processes.
The electromagnetic (photoproduction) process is, however, well known as 
a more ''clean'' process. 
Furthermore, photoproduction process provides an easier way to ''see'' 
the $\Theta^+$ which contains
an antiquark, since all required constituents are already present in
the initial state\cite{Karliner:2004gr}. Other processes, such as $e^+e^-$ and
${\bar p}p$ annihilations, would produce the strangeness-antistrangeness
from gluons, which has a consequence of the suppressed cross 
section\cite{Titov:2004wt}.

Several $\Theta^+$ photoproduction studies have been performed by using 
isobar models with Born approximation\cite{yu&ji2004}, where the
obtained cross section spans from several nanobarns to almost one $\mu$barn,
depending on the $\Theta^+$ width, parity, hadronic form factor cut-off, 
and the exchanged particles used in the process. Those parameters are 
unfortunately still uncertain at present. 

In this paper, we calculate the photoproduction cross sections by utilizing
an isobar model. Since the production threshold is already high we compare
the results with those obtained from a Regge model. The comparison is also
very important, since most input parameters in the isobar model are less
known.

\section{Formalism}
\begin{figure}[ht]
\centerline{\epsfxsize=4in\epsfbox{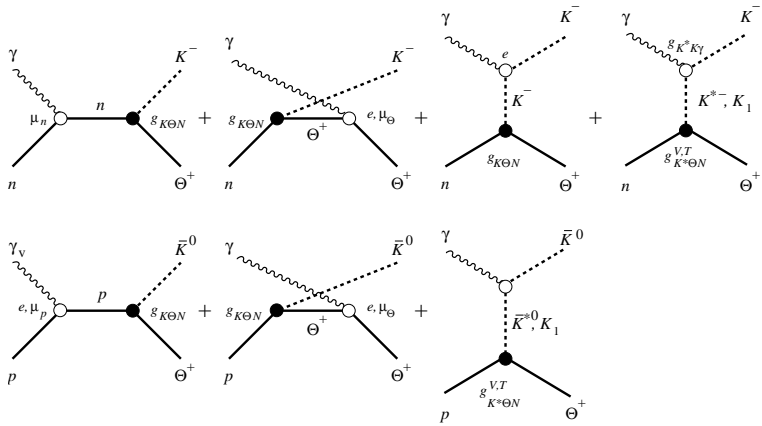}}   
\caption{Feynman diagrams for $\Theta^+$ photoproduction on neutron 
  $\gamma + n \longrightarrow K^- + \Theta^+$ (top)
  and on the proton $\gamma + p \longrightarrow {\bar K}^0 + \Theta^+$
  (bottom). \label{fig:feynman}}
\end{figure}

The basic background amplitudes for the processes 
\begin{eqnarray*}
\gamma (k) + n (p) \to K^- (q) + \Theta^+(p')  ~~\textrm{and}~~
\gamma (k) + p (p) \to {\bar K}^0 (q) + \Theta^+(p')
\end{eqnarray*}
are obtained from a series of tree-level Feynman diagrams shown in 
Fig.\,\ref{fig:feynman}. They contain the $n$, $\Theta^+$,  $K^-$, $K^{*-}$ 
and $K_1$ intermediate states 
in the first process, whereas in the second process the 
${\bar K}^0$ exchange does not present since a real photon cannot
interact with a neutral meson. The $K^*$ and $K_1$ intermediate states are considered
here, since previous studies on $K\Lambda$ and $K\Sigma$ photoproductions
have shown that their roles are significant.

The transition matrix for both reactions can be decomposed into 
\begin{eqnarray}
M_{\mathrm fi} &=& {\bar u}({\mbox{\boldmath ${p}$}}') 
  \sum_{i=1}^{4} A_i~M_i ~u({\mbox{\boldmath ${p}$}}) ~,
\label{eq:mfi}
\end{eqnarray}
where the gauge and Lorentz invariant matrices $M_i$ are given in, e.g.,
Ref.\cite{Lee:1999kd}.
In terms of Mandelstam variables $s$, $u$, and $t$, 
the functions $A_i$ are given by
\begin{eqnarray}
\label{eq:a1}
A_{1} & = & -\frac{e g_{\Theta}}{s - m_{N}^{2}} \left(Q_{N} +
\kappa_{N} \frac{m_{N} - m_{\Theta}}{2 m_{N}} \right) F_1(s) 
 - \frac{e g_{\Theta}}{u - m_{\Theta}^{2} + im_\Theta\Gamma_\Theta} \times
\nonumber\\&& 
 \left[ Q_{\Theta} + \kappa_{\Theta} \left(
\frac{m_{\Theta} - m_{N}}{2 m_{\Theta}} - i\,\frac{\Gamma_\Theta}{4m_\Theta}\right) 
\right] F_2(u) \nonumber\\
&& - \frac{C_{K^*}G^TF_3(t)}{M(t-m_{K^*}^2+im_{K^*}\Gamma_{K^*})(m_\Theta + m_N)} ~, \\
A_{2} & = & \frac{2e g_{\Theta}}{t - m_{K}^{2}} 
\left(\frac{Q_{N}}{s - m_{N}^{2}} + \frac{Q_{\Theta}}{u - m_{\Theta}^{2}} \right)
{\widetilde F}
+ \frac{C_{K^*}G^TF_3(t)}{M(t-m_{K^*}^2+im_{K^*}\Gamma_{K^*})} \nonumber\\&& 
\times \frac{1}{(m_\Theta + m_N)}
- \frac{C_{K_1}G^T_{K_1}F_3(t)}{M(t-m_{K_1}^2+im_{K_1}\Gamma_{K_1})(m_\Theta + m_p)}
,\label{eq:a2} \\
A_{3} & = & \frac{e g_{\Theta}}{s - m_{N}^{2}}~
\frac{\kappa_{N} F_1(s)}{2 m_{N}} - \frac{e g_{\Theta}}{u - 
m_{\Theta}^{2}}~ \frac{\kappa_{\Theta} F_2(u)}{2 m_{\Theta}} 
- \frac{C_{K^*}G^TF_3(t)}{M(t-m_{K^*}^2+im_{K^*}\Gamma_{K^*})}\nonumber\\&& 
\times \frac{m_\Theta - m_N}{m_\Theta + m_N} 
+ \frac{(m_\Theta +m_p)C_{K_1}G^V_{K_1}+ 
(m_\Theta -m_p)C_{K_1}G^T_{K_{1}}}{M(t-m_{K_1}^2+im_{K_1}\Gamma_{K_1})
(m_\Theta + m_p)}F_3(t)\label{eq:a3} \\
A_{4} & = & \frac{e g_{\Theta}\kappa_{N}}{s - m_{N}^{2}}
\frac{F_1(s)}{2 m_{N}} +
\frac{e g_{\Theta}\kappa_{\Theta}}{u - m_{\Theta}^{2}}~ 
\frac{F_2(u)}{2 m_{\Theta}} 
+ \frac{C_{K^*}G^VF_3(t)}{M(t-m_{K^*}^2+im_{K^*}\Gamma_{K^*})} ,
\label{eq:a4}
\end{eqnarray}
with $g_{\Theta}=g_{K \Theta N}$, 
$Q_\Theta =1$, $Q_N=1\, (0)$ for proton (neutron), $\kappa_N$ and $\kappa_\Theta$
indicate the anomalous magnetic moments of the nucleon and $\Theta^+$, and $M$
is taken to be 1 GeV in order to make the coupling constants
$ G^{V,T} = g^{V,T}_{K^*\Theta N}\, g_{K^* K\gamma}$ 
dimensionless.

The inclusion of hadronic form factors at hadronic vertices is performed
by utilizing the Haberzettl prescription\cite{Haberzettl:1998eq}. The form factors
are taken as
\begin{eqnarray}
  \label{eq:form_factor}
  F_i(q^2) &=& \frac{\Lambda^4}{\Lambda^4 + (q^2-m_i^2)^2} ~~~~~~
  \textrm{with} ~~~~~ q^2 ~=~ s,u,t ~,
\end{eqnarray}
with $\Lambda$ the corresponding cut-off.
The form factor for non-gauge-invariant terms ${\widetilde F}(s,u,t)$ 
in Eq.\,(\ref{eq:a2}) is extra constructed in order to satisfy crossing 
symmetry and to avoid a pole in the amplitude\cite{Davidson:2001rk}, i.e.,
\begin{eqnarray}
  \label{eq:fhat}
  \widetilde{F}(s,u,t) &=& F_1(s)+F_1(u)+F_3(t)-F_1(s)F_1(u) \nonumber\\&&
           -F_1(s)F_3(t) - F_1(u)F_3(t) 
           + F_1(s)F_1(u)F_3(t) .
\end{eqnarray}
Since $\Theta^+$ is an isoscalar particle, the coupling constants relations 
read
\begin{eqnarray}
  \label{eq:cc1}
   g_{K\Theta N} = g_{K^- \Theta^+ n} = g_{{\bar K}^0 \Theta^+ p} ~~,~~
   g^{V,T}_{K^*\Theta N} = g^{V,T}_{K^{*-} \Theta^+ n} = g^{V,T}_{{\bar K}^{*0} \Theta^+ p} ~.
\end{eqnarray}
The coupling constant $g_{K^- \Theta^+ n}$ can be calculated from the decay width of
the $\Theta^+\to K^+ n$ by using
\begin{eqnarray}
  \label{eq:width}
  \Gamma &=& \frac{g^2_{K^- \Theta^+ n}}{4\pi}\,\frac{E_n-m_n}{m_\Theta}\, p ~,
\end{eqnarray}
with $ p = [\{m_\Theta^2-(m_K+m_n)^2\}\{m_\Theta^2-(m_n-m_K)^2\}^{1/2}]/{2m_\Theta}$.
The precise measurement of the decay width is still lacking due to the
experimental resolution. The reported width\cite{experiment} is in the range of 
6--25 MeV. Using the partial wave 
analysis of $K^+N$ data Arndt {\it et al.}\cite{arndt}  found $\Gamma\le 1 $ MeV,
whereas the PDG\cite{pdg2004} announces 
$\Gamma = 0.9\pm 0.3 $ MeV. Based on this information we  use a width of
1 MeV in our calculation. Explicitly, we use 
\begin{eqnarray}
  \label{eq:cc_num}
  {g_{K\Theta N}}/{\sqrt{4\pi}} &=& 0.39 ~.
\end{eqnarray}
The magnetic moment of $\Theta^+$ is also not well known. A recent chiral soliton 
calculation\cite{kim2003} yields a value of $\mu_\Theta = 0.82 ~ \mu_N$,
from which we obtain $\kappa_\Theta=0.35$. Note that in the second channel 
the Regge model does not depend on this coupling constant as well as
the $\Theta^+$ magnetic moment.

The coefficient $C_{K^*}$ in Eqs.\,(\ref{eq:a1})-(\ref{eq:a4}) 
is introduced since in ${\bar K}^0$ photoproduction the vector meson exchange
in the $t$-channel is $K^{*0}$. The coefficient reads\cite{Mart:1995wu}
\begin{eqnarray}
  C_{K^*} &=& 
    1 
    {\rm ~~for~~} K^-\Theta^+ 
    ~~[-1.53 
    {\rm ~~for~~} {\bar K}^0\Theta^+]
    ~.
\end{eqnarray}

The coupling constants $g^{V}_{K^{*} \Theta N}$ and 
$g^{T}_{K^{*} \Theta N}$ are also not well known. Therefore, we follow 
Refs.\,\cite{yu&ji2004,liu&ko2004}, i.e., using $g^{V}_{K^{*} \Theta N}=1.32$
and neglecting $g^{T}_{K^{*} \Theta N}$ due to the lack of information
on this coupling. By combining the electromagnetic and hadronic coupling
constants we obtain
\begin{eqnarray}
  \label{eq:GVK*}
  {G^{V}_{K^*\Theta N}}/{4\pi} &=& 8.72\times 10^{-2} ~.
\end{eqnarray}

Most previous calculations excluded the $K_1$ exchange, mainly due to the
lack of information on the corresponding coupling constants. 
Reference\cite{yu&ji2004} used the vector dominance relation 
$g_{K_1K\gamma}=eg_{K_1K\rho}/f_\rho$ to determine the electromagnetic
coupling $g_{K_1K\gamma}$, where $f^2_\rho/4\pi=2.9$ and
$g_{K_1K\rho}=12$ is taken from the effective Lagrangian calculation
of Ref.\,\cite{haglin94}. As in the case of $K^*$, the $K_1$ hadronic 
tensor coupling will be neglected in this calculation due to the same 
reason. Following Ref.\,\cite{yu&ji2004}, the $K_1$ axial vector 
coupling $g^{V}_{K_1\Theta N}$ is estimated from an isobar model
for $K^+\Lambda$ photoproduction\cite{wjc} by using the extracted ratio 
$G^{V}_{K^*\Lambda N}/ G^{V}_{K_1\Lambda N}=-8.26$.
We note that the same ratio is also obtained in Ref.\,\cite{Mart:2000ed}
for the model without missing resonance $D_{13}(1895)$.
Therefore, in our calculation we use
\begin{eqnarray}
  \label{eq:GVK1}
  {G^{V}_{K_{1}\Theta N}}/{4\pi} &=& -7.64\times 10^{-3} ~.
\end{eqnarray}
The constant $C_{K_1}$ in Eqs.(\ref{eq:a2}) and (\ref{eq:a3}) is extracted
from fitting an isobar model to the $K^+\Sigma^0$ and $K^0\Sigma^+$ photoproduction
data\cite{Mart:jv}, i.e.,
\begin{eqnarray}
  \label{eq:ck1}
  C_{K_1} &=& 
    1 
    {\rm ~~for~~} K^-\Theta^+ 
    ~~[-0.17  {\rm ~~for~~} {\bar K}^0\Theta^+]
  ~.
\end{eqnarray}

\section{Regge Model}
In Regge model one should only use the $K^-$ and $K^*$ 
($K^*$ and $K_1$) diagrams in Fig.\,\ref{fig:feynman}
for the $\gamma n\to K^-\Theta^+$ ($\gamma p\to {\bar K}^0 \Theta^+$) 
channel.
Hence, the result from Regge model will not depend on the value
of $g_{K\Theta N}$ and $\Theta^+$ magnetic moment
in the second channel.
The procedure is adopted from Ref.\,\cite{guidal97}, i.e., by 
replacing the Feynman propagator with the Regge propagator
\begin{eqnarray}
  \label{eq:regge}
  P_{\rm Regge} &=& \frac{s^{\alpha_{K^i}(t)-1}}{\sin [\pi\alpha_{K^i}(t)]}
                    ~ e^{-i\pi\alpha_{K^i}(t)} ~ 
                    \frac{\pi\alpha_{K^i}'}{\Gamma [\pi\alpha_{K^i}(t)]} ~,
\end{eqnarray}
where $K^i$ refers to $K^*$ and $K_1$, and 
$\alpha_{K^i} (t) = \alpha_0 + \alpha '\, t$ denotes the corresponding
trajectory\cite{guidal97}. 

\section{Results and Discussion}

\begin{figure}[b]
\centerline{\epsfxsize=3.5in\epsfbox{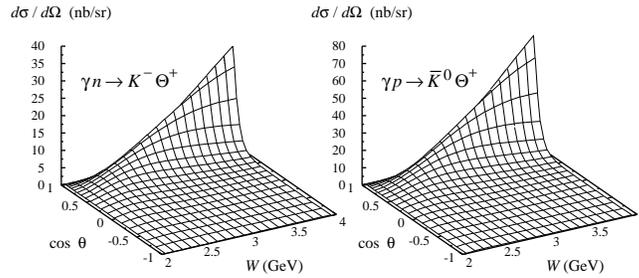}}   
\caption{Differential cross sections obtained by using the isobar model. \label{fig:d3dim}}
\end{figure}
The differential cross sections obtained from the isobar model in both channels
are shown in Fig.\,\ref{fig:d3dim}. Obviously, both channels show a forward
peaking differential cross section which is due to the strong contribution from
the $K^*$ intermediate state. Previous studies which use only Born 
terms\cite{yu&ji2004} obtained a backward peaking cross section for the
$\gamma p\to {\bar K}^0 \Theta^+$ channel, since in this case no
$t$-channel intermediate state is included. 
Figure\,\ref{fig:d3dim} also demonstrates
that the hadronic form factors are unable to suppress the cross sections
at higher energies.

\begin{figure}[ht]
\centerline{\epsfxsize=3.5in\epsfbox{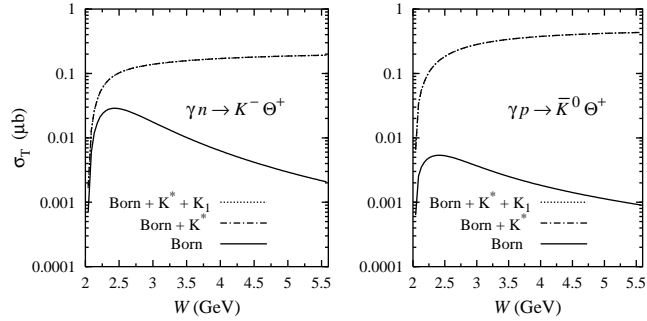}}   
\caption{Contribution of the Born terms, $K^*$- and $K_1$-exchange
  to the total cross sections. \label{fig:born}}
\end{figure}

The strong contribution of the $K^*$ in both
channels can be observed in Fig.\,\ref{fig:born}, where we can see that the
inclusion of this state increases the total cross sections by more than one
order of magnitude. In contrast to the $K^*$, contribution from the $K_1$
vector meson is negligible. This fact can be traced back to the
coupling constants given by Eqs.\,(\ref{eq:GVK1}) and (\ref{eq:ck1}). 

\begin{figure}[ht]
\centerline{\epsfxsize=3.5in\epsfbox{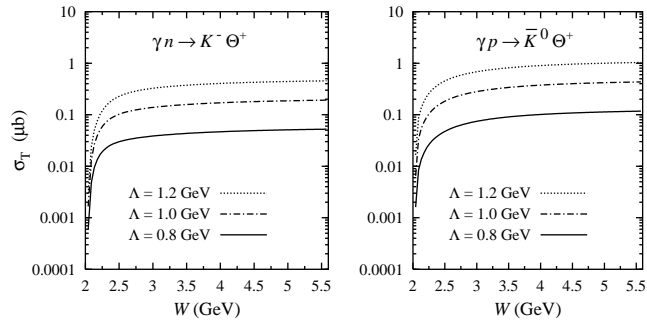}}   
\caption{Total cross sections for $\Theta^+$ photoproduction
  off a neutron (left) and a proton (right) as a function of the hadronic form factor
  cut-off $\Lambda$. \label{fig:parameter}}
\end{figure}

\begin{figure}[ht]
\centerline{\epsfxsize=3.2in\epsfbox{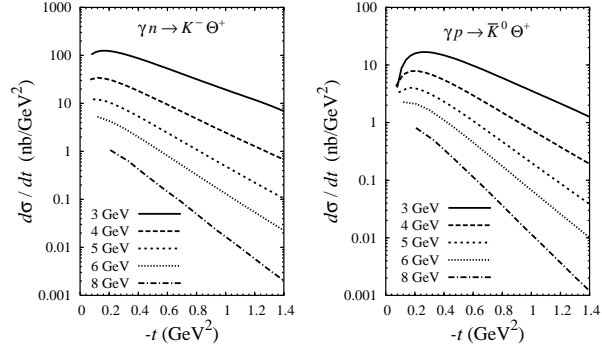}}   
\caption{Differential cross section for $\Theta^+$ photoproduction
  obtained from the Regge calculation. 
  The corresponding total c.m. energy $W$ is shown in each panel. \label{fig:regge}}
\end{figure}

\begin{figure}[ht]
\centerline{\epsfxsize=2in\epsfbox{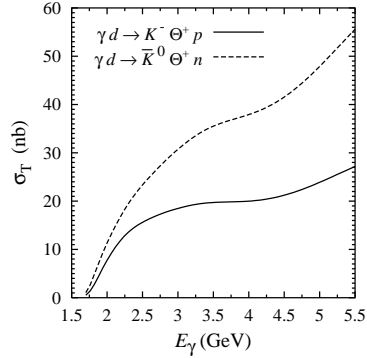}}   
\caption{Total cross section for the inclusive $\Theta^+$ photoproduction 
  on the deuteron. \label{fig:deuteron}}
\end{figure}

Figure \ref{fig:parameter} demonstrates the sensitivity of the total cross
sections to the choice of the hadronic form factor cut-off. Clearly, a right
choice of the cut-off is very important in this case. For this purpose, we
calculate also the cross sections by using a Regge model.
The results  are shown in 
Fig.\,\ref{fig:regge}. Obviously, the Regge approach predicts
smaller cross sections than those obtained from the isobar model.
In the case of $K\Lambda$ and $K\Sigma$ photoproductions, 
Ref.\cite{Mart:2003yb} showed that Regge model works nicely at
higher energies (up to $W=5$ GeV) but overpredicts the $K^+\Lambda$ 
(underpredicts the $K^+\Sigma^0$) data at the resonance region
($W\le 2$ GeV) by up to 50\%. Thus, we would expect the same result
for $\Theta^+$ photoproduction. By comparing with the result obtained
from the isobar model, we can conclude that the isobar prediction 
could overestimate the realistic cross section, especially at higher
energies, unless a softer hadronic form factor is chosen. 
This result can partly explain why the high energy experiments
are unable to observe the existence of the $\Theta^+$.

Using the elementary operator of the isobar model we predict the
inclusive total cross section for $\Theta^+$ photoproduction on the
deuteron. The results for both possible channels are given in 
Fig.\,\ref{fig:deuteron}, where we show the inclusive total
cross section obtained by using an isobar model with 
$\Lambda=0.8$ GeV. The fact that the $K^-\Theta^+$ cross section 
is smaller than the $K^0\Theta^+$ one is originated from the
elementary process (see Fig.\,\ref{fig:born}). 

In conclusion, we have calculated cross sections of $\Theta^+$
photoproduction by using an isobar and a Regge models. The Regge
model predicts smaller cross sections, especially at higher energies.

The work of TM has been partly supported by the QUE project.

\end{document}